\documentclass[aps,prl,twocolumn,letterpaper,superscriptaddress,showpacs]{revtex4-2}
\usepackage{graphicx}
\usepackage{CJK}
\usepackage{tikz}

\usepackage{verbatim}
\usepackage{lineno}
\usepackage{bm}
\usepackage{enumitem}
\usepackage{mathptmx}
\usepackage{subfigure}
\usepackage{amsmath}
\usepackage{physics}
\usepackage{subfiles}
\usepackage{natbib}
\usepackage{lipsum}
\usepackage{appendix}
\makeatletter

\usepackage{titlesec}
\titleformat*{\section}{\large\centering\bfseries}
\begin{document}


\title{Transport properties of a half-filled Chern band at the electron and composite fermion phases}

\date{\today}

\author{Ady Stern}
\email{adiel.stern@weizmann.ac.il}
\affiliation{\small{Department of Condensed Matter Physics\\
Weizmann Institute of Science, Rehovot 7610001, Israel}}

\author{Liang Fu}
\email{liangfu@mit.edu}
\affiliation{Department of Physics, Massachusetts Institute of Technology, Cambridge, MA 02139, USA}

\begin{abstract}

We consider a half-filled Chern band and its transport properties in two phases that it may form, the electronic Fermi liquid and the composite-fermion Fermi liquid. For weak disorder, we show that the Hall resistivity for the former phase is very small, while for the latter it is close to $2h/e^2$, independent of the distribution of the Berry curvature in the band. At rising temperature and high frequency, 
we expect the Hall resistivity of the electronic phase to rise, and that of the composite-fermion phase to deviate from $2h/e^2$. At high frequency, sign changes are expected as well. Considering high-frequency transport, we show that the composite fermion phase carries a gapped plasmon mode which does not originate from long ranged Coulomb interaction, and we show how this mode, together with the reflection of electro-magnetic waves off the Chern band, allow for a measurement of the composite-fermion Drude weight and Berry curvature. Finally, we consider a scenario of a mixed-phase transition between the two phases, for example as a function of displacement-field, and show that such transition involves an enhancement of the longitudinal resistivity, as observed  experimentally. 

\end{abstract}
                              




\maketitle

About a decade after the discovery of the fractional quantum Hall effect (FQHE), it was realized that interacting electrons in a strong magnetic field may form compressible metallic states with unique properties, which are not shared by conventional electron metals. Most prominent of these metallic states has been the half-filled Landau level, whose properties were understood by a flux-attachment mapping from electrons in a strong magnetic field to composite fermions at zero magnetic field\cite{PhysRevB.47.7312, PhysRevB.52.5890, halperin2020halffull, Jain2007}. 

Recently, the world of fractionalized states of matter was dramatically expanded with the experimental discovery of Fractional Chern Insulators (FCIs)  
that spontaneously break time reversal symmetry \cite{cai2023signatures, zeng2023thermodynamic,  park2023observation,xu2023observation, Lu_Han_Yao_Reddy_Yang_Seo_Watanabe_Taniguchi_Fu_Ju_2023}. FCIs are states  believed to be topologically identical to the FQHE, but formed in two-dimensional systems with Chern bands, without requiring a magnetic field \cite{tang2011high, sun2011nearly, neupert2011fractional, sheng2011fractional, regnault2011fractional, qi2011generic, Xiao_Zhu_Ran_Nagaosa_Okamoto_2011}.
Experimentally, FCIs with well quantized Hall conductivities at zero magnetic field were observed in twisted bilayer MoTe$_2$ \cite{park2023observation,xu2023observation} and pentalayer graphene/hBN heterostructure \cite{Lu_Han_Yao_Reddy_Yang_Seo_Watanabe_Taniguchi_Fu_Ju_2023}. The role of the Landau level filling in the FQHE was played by the filling of  moir\'e bands with Chern number $C= \pm 1$ \cite{wu2019topological, devakul2021magic,  li2021spontaneous, crepel2023anomalous, Park_Kim_Chittari_Jung_2023, Dong_Patri_Senthil_2023, Dong_Wang_Wang_Soejima_Zaletel_Vishwanath_Parker_2023, Zhou_Yang_Zhang_2023}. Furthermore, a compressible state was measured at a half filled-Chern band.

Hand in hand with the similarities, there are important differences between the FQHE and the FCI states. Prominent among these are the energy scales. For a clean FQHE state there are two energy scales, the electron-electron interaction energy $e^2/l_B$ and the cyclotron energy $\hbar\omega_c$ (here $e$ is the electron charge, $l_B$ the magnetic length and $\omega_c$ the cyclotron frequency). The incompressible FQHE and compressible composite fermion ground states form at an arbitrarily small electron-electron interaction. The Chern band generally has an energy width, which introduces another energy scale. As a consequence, for arbitrarily small interaction, it is sensible to expect the electrons in a partially filled Chern band to form a Fermi liquid, for any fractional band filling, such that the FCI forms when the interaction passes a threshold scale, above which the band filling becomes a crucial parameter. 

In this work, we consider compressible metallic states in a half filled Chern band for weak interaction (Fermi liquid of electrons) and for strong interactions (Fermi liquid of composite fermions)\cite{dong_composite_2023, goldman2023zero}. Since the strength  of the interaction is determined with reference to the band structure, a transition between the limits may be induced by knobs that affect the latter, such as a displacement field. We focus on properties which are independent of the details of the half-filled Chern band. We start by formulating the composite fermion picture. Then, we analyze the $dc$ and $ac$ conductivity and resistivity matrices of both states, as well as their experimental implications and their temperature dependence. We find that a crucial difference between FQHE and FCI composite fermions emerges from the anomalous velocity of the latter. Finally, we analyze transport in the transition region between the electron Fermi liquid and the composite fermion fermi liquid. 

{\noindent \it Composite fermion theory:} At the half filled Landau level \cite{PhysRevB.47.7312, PhysRevB.52.5890, halperin2020halffull}, electrons are transformed into composite fermions by a singular gauge transformation that attaches two flux quanta to each electron. The transformed Hamiltonian is that of composite fermions interacting with a static, externally applied, magnetic field and a dynamical magnetic field generated by these flux quanta. At the mean field level the dynamical field is approximated by its expectation value with respect to a state of uniform density, such that the two magnetic fields cancel one another, and the composite fermions form a zero-field Fermi liquid. The theory has been very successful in describing some states of the $\nu=N+1/2$ series, e.g., $N=0,1$ for electrons in a quadratic dispersion. Its extension to include Cooper pairing of composite fermions has been successful in describing the $N=2,3$ cases. It has been, however, inadequate in describing states of $N>3$, in which translation symmetry is spontaneously broken. 

Here we consider  a half filled Chern band with a Berry curvature $F({\bf k})$. We employ a continuum model for the band as appropriate for the observed FCIs in moir\'e materials, as opposed to the tight-binding models employed in earlier formulations\cite{PhysRevB.97.125131}. The Hamiltonian then takes the form 
\begin{equation}
    H=H_0(p, r)+H_{int}
    \label{Hscheme}
\end{equation}
with 
$H_0(p, r)$ being 
the non-interacting Hamiltonian and $H_{int}$ describing the interaction between the electrons. 
We assume that the noninteracting Hamiltonian $H_0$ (for spin/valley polarized electrons) already breaks time-reversal symmetry and produces Chern bands, one of which is half-filled.   
We perform flux attachment in the same way it is done in the FQHE context, which modifies the kinetic part of the Hamiltonian by $p\rightarrow p-a(r)$, with $\nabla\times a(r)= 2\Phi_0 n(r)$, attachine two flux quanta to each electron. At the level of mean field theory, for  a half filled band, $n(r)$ corresponds to half an electron per unit cell, and therefore the attached flux corresponds to one flux quantum per unit cell. WIth an integer number of flux quanta per unite cell, 
the unit cell does not change, and therefore the number of bands does not change, too.  The transformation then maps a half filled band of electrons onto a half filled band of composite fermions. Importantly, time reversal symmetry is explicitly broken by the Chern band Hamiltonian $H_0$ as well as the  flux attachment $\nabla \times a(r)$. As a consequence, the half filled band of composite fermions may have a Berry curvature, and an anomalous Hall conductivity. 

Just as composite fermion theory was successful in analyzing the half filled Landau levels for some values of $N$ and not for others, we expect it to be successful in describing some half filled Chern bands, and not others. We now give several examples in which the theory is likely to be successful. 

The first example is that of a Twisted Bi-layer Graphene at the magic angle in the chiral limit \cite{tarnopolsky2019origin}. In this system  each valley carries two perfectly  flat bands, which are sublattice-polarized and carry Chern numebrs of $C=\pm1$. Spontaneous breaking of time reversal symmetry leads to filling of one of the two bands, say the $C=+1$, with the other one being pushed up in energy by an exchange gap. 
Each of the two bands have one state per unit cell, and they are separated by an energy gap from the adjacent dispersive bands. At the energy window between the two flat bands the system's Chern number is $+1$. As a magnetic flux $\tilde\phi$ per electron is turned on, the bands do not disperse, but states are transferred between them, making the number of states per unit cell in the two bands $1\pm\frac{\tilde\phi}{2}$ \cite{PhysRevB.104.L121405}.  We turn on a flux that empties the $C=+1$ band, that is, a magnetic  field that is anti-parallel to $F$. At the point ${\tilde\phi}=2$ the $C=+1$ band becomes empty, and the Chern number changes by one. Such a change must be accompanied by a closure of the energy gap between the flat and dispersive bands\cite{PhysRevB.104.L121405}. Thus, the flux attachment maps the problem of a half filled perfectly flat Chern band of electrons to the problem of composite fermions in a partially filled dispersive band. In view of the dispersion of the band, a Fermi liquid of composite fermions becomes a sensible starting point. We note that the composite fermions band has no time reversal symmetry, and is therefore likely to have a Berry curvature.

In fact, the perfectly flat bands in the chiral limit are spanned by wave functions which are closely related to those of the lowest Landau level of electrons in a uniform magnetic field  $B=\Phi_0/a^2$, augmented by a periodic magnetic field, making the relation to the FQHE composite fermion theory more transparent. Furthermore, recent numerical and theoretical work suggests criteria for the suitability of Chern bands to the formation of FCI states \cite{wang2021exact, ledwith2022vortexability}, and these criteria have been shown to be satisfied by bands composed of lowest Landau levels on a curved space\cite{PhysRevResearch.5.L032048}. Again, the correspondence to the lowest Landau level suggests the suitability of composite fermion states to form. 

Finally, the composite Fermi liquid state has been found at half filling of the Chern band in recent numerical studies of twisted bilayer MoTe$_2$ \cite{goldman2023zero,dong_composite_2023}. In this system, interlayer tunneling and intralayer moir\'e potential combine to produce a spatially-varying ``Zeeman'' field with the periodicity of the moir\'e superlattice, which couples to electron's layer pseudospin \cite{wu2019topological}. Importantly, the pseudospin texture in real space is topologically nontrivial and induces a pseudomagnetic field with one flux quantum per unit cell. 
When the pseudomagnetic field is nearly uniform, the resulting bands strongly resemble Landau levels \cite{paul2023giant, morales2023magic}, thus giving rise to FCIs and composite Fermi liquids at fractional band fillings.    
In particular,  at half filling, the attachment of $\tilde\phi=2$ flux quanta to electrons is energetically preferred, because this flux attachment cancels the pseudomagnetic field {\it on average}, leading to zero {\it net} field for composite fermions. However, it seems unlikely that the effective field experienced by composite fermions is everywhere zero. Most likely, there remains an effective field that is spatially varying and has zero net flux per unit cell, which breaks time-reversal symmetry in the composite fermion mean-field theory. Then, the problem of half-filled Chern band in twisted bilayer MoTe$_2$ is mapped to the problem of composite fermions in a time-reversal-breaking band which is expected to carry non-zero Berry curvature, thus giving rise to anomalous Hall effect of composite fermions.      
 


We now assume that the half filled Chern band is such that for a sufficiently strong interaction a composite fermion state is formed, and analyze its properties, contrasting them with those of the half filled Landau level, and the half filled Chern band at weak interaction. 

{\noindent \it DC transport} 
For non-interacting or weakly interacting electrons, linear transport in partially filled bands with Berry curvature is most easily analyzed by means of a Boltzmann equation, in which the occupation function $f({\bf r},{\bf k},t)$ is calculated. Since the anomalous velocity $v_H(k)=e{\bf E}\times{\bf F(k)}$ is linear in the electric field ${\bf E}$, its contribution to the current may be calculated by summing over the states that are occupied in equilibrium, in which $f=f_0({\bf k})\equiv \Theta(\mu-E({\bf k})$, with $\mu$ the chemical potential. The longitudinal and Hall conductivities are then, 
\begin{eqnarray}
\sigma_{xx}&=&e^2\int d\theta\nu(\theta){\bf v}_x(\theta)\cdot {\bf v_x}(\theta=0)\tau({\theta})\equiv\frac{e^2}{h}{\cal D}\tau \\
\sigma_{xy}^F&=&\int d{\bf k}F({\bf k})f_0({\bf k})
\label{BoltzmannConductivities}
\end{eqnarray}
 where $\nu(\theta)$ is the density of states at the Fermi energy in an angle $\theta$ with respect to the $x$-axis, ${\bf v}_x(\theta)$ is the $x$-component of the velocity, $\tau(\theta)$ is the mean free time, within the relaxation time approximation, and $\cal D$ is the Drude weight, defined such that it carries the units of frequency.   

The conductivity of the electronic Fermi liquid is then, 
\begin{equation}
    \sigma_{ij}=\sigma_{xx}\delta_{ij}+\sigma_{xy}^F\epsilon_{ij}
    \label{efl-conductivity}
\end{equation}
For very clean systems with a partially-filled conduction band the $\omega=0$ conductivity matrix of non-interacting electrons is dominated by the diagonal matrix elements. Consequently, the Hall resistivity is $\propto 1/({\cal D}\tau)^2$, and is much smaller than the longitudinal one, which is $1/({\cal D}\tau)$.  The Hall angle is small, and $\rho_{xy}$ is non-universal and depends on relaxation time $\tau$.    

 In contrast to the electrons' Fermi liquid case in which two conductivities were added (see Eq. (\ref{efl-conductivity})), in composite fermion theory two resistivities are added to obtain the electrical resistivity - the resistivity $\sigma_{{\rm  cf}}^{-1}$ of a composite fermion liquid and the Chern-Simons resistivity, $\rho_{\rm cs}=2h/e^2 \epsilon_{ij}$, 
\begin{equation}
    \rho^e=\sigma_{\rm cf}^{-1}+\rho_{\rm cs}
    \label{CFtoE}
\end{equation}

For the half-filled Landau level, $\sigma_{\rm cf}$,  the conductivity matrix of fermions at zero magnetic field, is diagonal. Consequently, within mean-field theory $\rho_{xy}=2h/e^2$ at $\nu=1/2$, independent of the longitudinal response of the composite fermions. Our case is different, since the Composite fermion conductivity matrix has also off-diagonal components,  $\sigma_{ij}^{\rm cf}=({\cal D}^{\rm cf}\tau^{\rm cf})\delta_{ij}+\sigma_{xy}^{\rm cf}\epsilon_{ij}$, due to the Berry curvature in the composite fermion band. As long as this matrix is dominated by its diagonal terms, the electrical resistivity matrix  will be 
\begin{equation}
    \rho^e\approx \frac{1}{{\cal D}^{\rm cf}\tau^{\rm cf}}\delta_{ij}+2\epsilon_{ij},
    \label{CFL-resistivity}
\end{equation}
neglecting terms of order $1/[({\cal D}^{\rm cf}\tau^{\rm cf}]^2$ and smaller. Importantly, in that limit the Hall resistivity is independent of the composite fermion Hall conductivity, i.e., of the details of their band structure. These will come to play later on, when we discuss cases where the composite fermions resistivity is not dominated by its diagonal terms. 

The resistivity matrices predicted for the electronic and composite fermions phases at the half-filled Chern bands are profoundly different even for perfectly clean systems. This should be contrasted with the half-filled Landau level, in which Galilean invariance dictates  the  Hall resistivity to be $\rho_{xy} = 2 h/e^2$. 
Interesting differences emerge also away from this limit. We consider three axes where that happens - increasing temperature, high frequency and the phase transition between the electronic and composite fermion liquid. 

{\noindent\it Temperature dependence of the resistivity}
For the electronic Fermi liquid, as the temperature increases ${\cal D}\tau$ decreases due to Umklapp scattering. As long as the temperature is much lower than the band width and the ferromagnetic ordering temperature, the Hall conductivity $\sigma^F$ due to the Berry curvature effect should be approximately temperature independent. Consequently, we expect $\rho_{xy}\approx \sigma^F/({\cal D}\tau)^2$ to increase with increasing temperature.   

For a temperature that is of the order the band width, all states of the band are approximately half-occupied, and $\sigma^F\approx \frac{1}{2}$. The longitudinal conductivity is likely to become much smaller than this, making the conductivity and resistivity matrices dominated by their off diagonal elements. At most, then, the Hall resistivity may get to a value of $2$. 

For the composite fermion liquid, when the temperature increases, ${\cal D}^{\rm cf}\tau^{\rm cf}$ is suppressed, making the Hall resistivity of the composite fermion significantly different from zero. When that happens, the electrical Hall resistivity significantly deviates from $2$. The sign of the deviation depends on the sign of the Berry curvature within the composite fermion band, which may, in principle, have any value. By the same arggument as above, for a temperature comparable to the band width, the Hall conductivity of the composite fermion band becomes $C_{\rm cf}/2$, with $C_{\rm cf}$ the integer Chern number of the composite fermion band. This leads to an electronic Hall resistivity of $2$ for $C_{\rm cf}=0$, and $2-\frac{2}{C_{\rm cf}}$ for $C_{\rm cf}\ne 0$.


{\noindent\it High-frequency transport}
The electronic and composite Fermi liquid phases are expected to  differ also in the frequency dependence of the Hall resistivity. In the electronic Fermi liquid, as $\omega$ increases above $1/\tau$, the longitudinal conductivity $\sigma_{xx}$ approaches ${\cal D}/i\omega$. It becomes imaginary, and decreases significantly in magnitude, whereas $\sigma_{xy}$ due to the Berry curvature is approximately frequency independent. As long as the conductivity matrix is dominated by its diagonal elemnts, $\rho_{xy}\approx -\sigma_{xy}/\sigma_{xx}^2$ increases with  frequency relative to its $dc$ value, and acquires an opposite sign. By the same argument, the composite fermion Hall resistivity $\rho^{\rm cf}_{xy}$ also increases with frequency, and flips its sign. However, as Eq.(\ref{CFtoE}) shows, the sign of the deviation of the electronic $\rho_{xy}$ from $2h/e^2$ depends on the relative sign between the composite fermions and Chern-Simons contributions. 

At high enough frequency, the diagonal and off-diagonal elements of the conductivity are comparable in magnitude.  Generally speaking, at frequencies for which optical absorption does not occur,  the conductivity of a two-dimensional isotropic system takes the general form 
\begin{equation} 
\sigma_{ij} (\omega) =  - i \sigma_{xx}''(\omega) \delta_{ij}+ \sigma'_{xy}(\omega) \epsilon_{ij}. 
\end{equation}
where $\sigma_{xx}''$ is the imaginary part of the longitudinal $ac$ conductivity and $\sigma'_{xy}$ is the real part of the $ac$ Hall conductivity, both of which are non-dissipative. This electrical conductivity matrix can have a zero eigenvalue, if at certain frequency $\omega_F$ we have  
$\sigma_{xx}''(\omega_F)=\sigma'_{xy}(\omega_F)$. Below, we explore this possibility for the half-filled Chern band.

A zero eigenvalue of the conductivity has an observable consequence on the reflection of electromagnetic waves by the two-dimensional system. When an electromagnetic wave  of frequency $\omega$ is irradiated along the $z$-axis on the two dimensional system, with the electric field of amplitude $E_0$, the amplitudes of the reflected and transmitted waves $E_r,E_t$ are 
\begin{eqnarray}
    E_r&=-&\frac{1}{1+{\alpha\sigma}}{\alpha\sigma} E_0 \\
    E_t&=&\frac{1}{1+{\alpha\sigma}}E_0
    \label{reflection}
     \end{eqnarray}
     where $\alpha$ is the fine structure constant \cite{liu_anomalous_2020, tse_magneto-optical_2011}. 
     
          At $\omega_F$, for $E_0$ linearly polarized in the $x$-direction, the reflected field is
\begin{equation}
    E_r=\frac{\alpha\sigma_{xy}}{1 - 2i \alpha\sigma_{xy}}E_0(i,1)
    \label{circularwave}
\end{equation}
The linearly polarized incident wave is then turned into a circularly polarized reflected wave. Equivalently, at normal incidence, circularly polarized light of one handedness is fully transmitted, whereas light of the opposite handedness is partly transmitted and partly reflected. Therefore, we call this phenomenon the ``chiral mirror'' effect. 

For the electronic Fermi liquid, at high frequency $\omega \gg 1/\tau$, the conductivity takes the form
\begin{equation} 
\sigma_{ij}=\frac{\cal D}{i\omega}\delta_{ij}+\sigma_{xy}^F\epsilon_{ij}. 
\label{efl-highfreq}
\end{equation}
with $\sigma_{xy}^F$ given in Eq. (\ref{BoltzmannConductivities}). The conductivity has a zero eigenvalue at $\omega_F=\pm {\cal D}/\sigma_{xy}^F$. In this case, the reflected amplitude measures the Hall conductivity $\sigma_{xy}^F$, while the frequency $\omega_F$ measures ${\cal D}/\sigma_{xy}^F$. 

For the composite fermion liquid, the composite fermion conductivity $\sigma_{\rm cf}$ takes a similar form as Eq.(\ref{efl-highfreq}), albeit with ${\cal D}_{\rm cf}$ and $\sigma_{xy}^{\rm cf}$ of composite fermions.  The electromagnetic response is determined by the electrical resistivity matrix shown in Eq.(\ref{CFtoE}). 
Employing Eq. (\ref{reflection}) again, we see that the frequency $\omega_F$ at which the electronic conductivity has a zero eigenvalue is the same frequency at which the composite fermion conductivity has a zero eigenvalue.   This is so since the composite fermions resistivity diverges at that frequency, and the addition of the Chern-Simons resistivity does not affect it. Thus, in this case an electronic response function directly reflects a composite fermion property.  At that frequency, again, a linearly polarized light is reflected as a circularly polarized one, but  Eq. (\ref{circularwave}) is replaced by 
\begin{equation}
    E_r=E_0 \frac{\frac{\alpha\sigma_{xy}^{\rm cf}}{4\sigma_{xy}^{\rm cf}-1}}{1-2i\frac{\alpha\sigma_{xy}^{\rm cf}}{4\sigma_{xy}^{\rm cf}-1}}(i,1)
    \label{circularwavecf}
\end{equation}

This phenomenon is unique to composite fermions in a Chern band, and does not occur for the FQHE, in which the composite fermion are not part of a band with Berry curvature. While the reflection is of order $\alpha$ for the electronic phase, it has the potential of being much larger for the composite fermion phase, if $\sigma_{xy}^{\rm cf}$ becomes close to $1/4$. Note that the frequencies discussed here are in the intra-band regime, or within the gap between bands. In this regime there is no inter-band absorption, and the response is dominated by virtual transitions of composite fermions. This is in contrast to the inter-band absorption discussed in [\onlinecite{dong_composite_2023}]. 

The high frequency response of the electronic and composite fermion phases differ in another experimentally observable way - the spectrum of the collective plasma excitations, whose long wave-length dispersion is the line $\omega(q)$ at which $\det{(1+iq^2V(q)\sigma(q,\omega)/\omega}=0$, with $V(q)$ being the Coulomb interaction. Alternatively, this condition may be written as $\det{\rho(q,\omega)+iq^2V(q)/\omega}=0$. Since $q^2V(q)\rightarrow 0$ as $q\rightarrow 0$, the $q\rightarrow 0$ plasma frequency occurs at the point where the determinant of the resistivity vanishes, in contrast to Eq. (\ref{circularwavecf}), which required the determinant of the conductivity to vanish. 

For an electronic Fermi liquid in a Chern band, the conductivity matrix (\ref{efl-highfreq}) leads to a gapless plasma frequency $\omega_p=\sqrt{{\cal D}q^2V(q)}$. In the absence of a screening gate $V(q)=2\pi e^2/q$, while a screening gate removes the $q\rightarrow 0$ divergence of $V(q)$. In contrast, for a composite fermion state the resistivity matrix (\ref{CFL-resistivity}) leads to a gapped plasma mode following $\omega(q)=\sqrt{\omega_0^2+\omega_p^2}$. This gapped plasmon is similar to Kohn's mode of electrons in a magnetic field, which follows 
$\omega=\sqrt{\omega_c^2+nq^2V(q)/m}$, with $\omega_c$ the cyclotron frequency. Remarkably, this is a Kohn mode with no magnetic field present, and no inter-Landau level gap. {\it Electron-electron interaction by itself leads to a gapped collective mode at $q=0$}. The frequency at which the mode emerges at $q=0$ is $\omega_0=2{\cal D}^{\rm cf}/|2\sigma_{xy}^{\rm cf}-1|$. An observation of the two high frequency phenomena we discusses here - the {\it chiral} reflection of electro-magnetic waves and the "cyclotron"-resonance of the composite fermions - would then extract both the anomalous Hall conductivity and the Drude weight of the composite fermions. 

At low $\omega$ and large $q$ the electronic conductivity of the composite fermion state should be linear in $q$ \cite{dong_composite_2023}, as is the case for the half filled Landau level\cite{PhysRevB.47.7312}. 


{\it Drude weight}  The Drude weight is a key quantity that governs low-frequency transport in metallic states.  In clean isotropic systems, the longitudinal conductivity $\sigma_{xx}$ at frequencies below interband transitions take the general form
\begin{eqnarray}
\sigma_{xx}(\omega) = {\cal D} \left( \delta(\omega) + \frac{i}{\pi \omega} \right).  
\end{eqnarray}
The effect of impurity scattering can be included phenomenologically in the Drude term by adopting a single-relaxation-time approximation, leading to the replacement of the term in the brackets by a Lorentian term with a width of $1/\tau$. 

As shown by Kohn \cite{Kohn1964}, the Drude weight is directly related to a thermodynamical property, the change of ground state energy $E_{0}$ with respect to a phase twist $\theta$ in the periodic boundary conditions to which the system is subjected. In two dimensions, $\cal D$ can be expressed as  
\begin{eqnarray}
{\cal D}=\frac{\pi}{h } \frac{\partial^2 E_{0}}{\partial^2 \theta}.  
\end{eqnarray}
where a factor of $e^2/h$ has been taken off the original definition. 
In free-electron systems with parabolic dispersion, the Drude weight ${\cal D} = n/m$ is simply determined by the charge density. On the other hand, under a magnetic field, the Drude weight $\cal D$ vanishes because the Landau level is completely dispersionless. 

For Chern bands with a non-zero bandwidth, in the absence of interactions, the ground state at partial band filling is a metal with a Fermi surface, and the Drude weight is nonzero. For example, the Drude weight in Landau levels perturbed by periodic potential is numerically calculated in Ref.\cite{Paul_Crowley_Fu_2023}. In contrast, the resistivity matrix (\ref{CFL-resistivity}) shows that the Drude weight of the electrons in the composite fermion phase vanishes. 
At present, we have no theoretical understanding of the transition between the electron and composite fermion states we analyze for the half filled Chern band. In a first order transition, $\cal D$ is discontinuous through the transition, but a scenario of a continuous transition, in which $\cal D$ varies continuously, is possible\cite{PhysRevB.86.075136}. For both scenarios, it is plausible that at the transition the system would be composed of a mixed phase, whose properties we now analyze.

{\noindent \it Transport in a mixed phase system} We consider a system made of a mixture of a Fermi liquid of composite fermions and a Fermi liquid of electrons. In general, for two-dimensional systems composed of two phases characterized by a local Ohm's law relation, the elements of the macroscopic conductivity and resistivity matrices  $\sigma^{mac},\rho^{mac}$ lie on a semi-circle\cite{dykhne_theory_1994, stern_strong_2002}. For the resistivity, denoting the resistivity matrices of the two phases by $\rho^{(1)}, \rho^{(2)}$, the semi-circle is, 
\begin{equation}
\left( \rho _{xy}-\bar{\rho}_{xy}\right) ^{2}+\rho _{xx}^{2}=\bar{\rho}%
_{xx}^{2}
\end{equation}
The values of $\bar{\rho}_{xx},\bar{\rho}_{xy}$ are determined by the
longitudinal and Hall resistivities of the two phases:
\begin{eqnarray}
\bar{\rho}_{xy} &=&\frac{1}{2}\frac{\det \rho ^{(2)}-\det \rho ^{(1)}}{\rho
_{xy}^{(2)}-\rho _{xy}^{(1)}}  \nonumber \\
\bar{\rho}_{xx} &=&\sqrt{\bar{\rho}_{xy}^{2}+\frac{\det \rho ^{(2)}\rho
_{xy}^{(1)}-\det \rho ^{(1)}\rho _{xy}^{(2)}}{\rho _{xy}^{(1)}-\rho
_{xy}^{(2)}}}  \label{semi-circ-general}
\end{eqnarray}
The position of the macroscopic resistivity on the semi-circle is determined by the relative fraction of the two phases, and lie between the two points that correspond to $\rho^{(1)}$ and $\rho^{(2)}$. The conductivity satisfies a similar relation, with each $\rho$ replaced by $\sigma$. 

Qualitatively, these relations imply that when the two phases are clean enough such that their longitudinal resistivities $\rho_{xx}^{(1)},\rho_{xx}^{(2)}$ are smaller than the difference between their Hall resistivities $|\rho_{xy}^{(1)}-\rho_{xy}^{(2)}|$, the transition between them involves a peak in the macroscopic longitudinal resistivity, whose maximum value is approximately $\frac{1}{2}|\rho_{xy}^{(1)}-\rho_{xy}^{(2)}|$. This is the case here, and this peak was indeed observed in experiment on pentalayer graphene/hBN \cite{Lu_Han_Yao_Reddy_Yang_Seo_Watanabe_Taniguchi_Fu_Ju_2023}. As seen in Eqs. (\ref{CFL-resistivity}) and (\ref{efl-conductivity}), in the clean limit $\rho_{xx}$ is very small in both phases, while $\rho_{xy}\approx 2$ for the composite fermion state, and $\rho_{xy}\approx 0$ for the electron state. In that limit $\bar{\rho}_{xy}=\bar{\rho}_{xx}\approx 1$. 

The semi-circle obtained for the conductivity is very different, since the longitudinal conductivity of the electronic state is very large. Consequently, $\bar{\sigma}_{xy}$ and $\bar{\sigma}_{xx} $ are both very large and the transition between the two conductivity values of the two phases is monotonic. 

To summarize, we explored in this paper the difference between transport properties of two possible states of a half-filled Chern band, the electron Fermi liquid and the composite fermion Fermi liquid. We presented predictions for unique properties of their high frequency linear response, for their $dc$ linear response and in particular its temperature dependence and for the transition region between the two states. 

\begin{acknowledgements}
It is our pleasure to thank Long Ju, Xiaodong Xu, Tonghang Han and Zhengguang Lu for stimulating discussions about experiments. We also thank Yuval Oreg, Omri Lesser, Felix von Oppen, Nisarga Paul and Philip Crowley for helpful discussions.  
AS was supported by grants from the ERC
under the European Union’s Horizon 2020 research and innovation programme
(grant agreements LEGOTOP No. 788715 and HQMAT No. 817799), the DFG
(CRC/Transregio 183, EI 519/7-1) and the ISF
Quantum Science and Technology (2074/19). LF was supported by the Simons Investigator Award from the Simons Foundation. 
\end{acknowledgements}

\section{Appendix: Reflection of an electromagnetic wave from a two dimensional system}

We consider an electromagnetic wave incident from along the $z$-axis on the half-filled Chern band lying on the $x-y$ plane. The amplitudes of the incident, transmitted and reflected electric field are ${\bf E}_0,{\bf E}_t,{\bf E_r}$ respectively. The wave-vectors are parallel to the $z$-axis, with the reflected wave-vector opposite in direction to the other two. The two-dimensional half-filled Chern band imposes boundary conditions at $z=0$, which are 
\begin{eqnarray}
    {\bf E}(z=0^+)&=&{\bf E}(z=0^-)\\
    {\bf B}(z=0^+)-{\bf B}(z=0^-)&=&\frac{4\pi}{c}{\bf J \times  {\hat z}}
    \label{interface}
\end{eqnarray}
where $\bf J$ is the current density, and $\bf E,B$ are the electric and magnetic fields. 

We write the electric field as ${\bf E}(z, t)={\bf E}(z) e^{-i\omega t}$ with   
\begin{eqnarray}
{\bf E}(z<0) &=& \left( {\bf E}_0 e^{i qz} + {\bf E}_r e^{-i qz}  \right), \nonumber \\
{\bf E}(z>0)&=& {\bf E}_t e^{i qz} 
\end{eqnarray}
Then, the magnetic field ${\bf B} = \frac{1}{i\omega} {\nabla \times \bf E}$ at $z=0^+$ and $0^-$ is given by
\begin{eqnarray}
{\bf B}(0^-) &=& \frac{1}{c} {\bf \hat{z}} \times \left( {\bf E}_0 - {\bf E}_r \right), \nonumber \\
{\bf B}(0^+)&=& \frac{1}{c} {\bf \hat{z}} \times {\bf E}_t  
\end{eqnarray}
The boundary conditions translate to 
\begin{eqnarray}
    {\bf E}_0+{\bf E}_r&=& {\bf E}_t \\
    {\bf E}_0- {\bf E}_r&=& (1+ 2 \alpha \sigma ) {\bf E}_t 
\end{eqnarray}
where $\sigma$ is the $2\times 2$ conductivity matrix in unit of $e^2/h$ and $\alpha$ is the fine structure constant. 
The solution is
\begin{eqnarray}
{\bf E}_r 
=-\frac{1}{1+ \alpha \sigma } \alpha \sigma {\bf E}_0
\end{eqnarray}


This derivation is a compact form of the one introduced in \cite{liu_anomalous_2020, tse_magneto-optical_2011}.

\bibliography{ref.bib}

\end{document}